$Ta_2NiSe_5$: a candidate topological excitonic insulator with multiple band inversions

Xiaobo Ma[#], Guangwei Wang[#], Huican Mao, Zhihong Yuan, Tianye Yu, Rui Liu, Yiran Peng, Pengyu Zheng, and Zhiping Yin*

*Department of Physics and Center for Advanced Quantum Studies, Beijing Normal University, Beijing 100875, China*

The electronic structures and topological properties of the orthorhombic and monoclinic phases of the quasi-one-dimensional excitonic insulator $Ta_2NiSe_5$ are investigated based on density functional theory. In contrast to a single parity or band inversion across the Fermi level in many topological insulators studied previously, there are multiple parity and band inversions with or without spin-orbit coupling in both phases of $Ta_2NiSe_5$, resulting in more complex and topologically nontrivial electronic structures. The Dirac cone type surface states of the low-temperature monoclinic phase are also obtained. In this paper, we demonstrate that $Ta_2NiSe_5$ is a promising candidate as a three-dimensional topological excitonic insulator.

## I. INTRODUCTION

Topological insulators (TIs) [1] have attracted considerable interest in condensed-matter physics [2-7] due to their promising application in spintronics and quantum computations [1,8]. TIs have insulating band gaps in the bulk, but gapless surface states that reside in the bulk band gaps [1], which are protected by time-reversal symmetry (TRS). Three-dimensional (3D) TIs with TRS are classified into strong and weak TIs (STIs and WTIs) characterized by four $Z_2$ invariants ($\nu_0$; $\nu_1$, $\nu_2$, $\nu_3$) [4,5]. Systems with $\nu_0 = 1$ are STIs. Systems in which $\nu_0 = 0$ and $\nu_1$, $\nu_2$, and $\nu_3$ are not all zero are WTIs. STIs and WTIs have topological surface states consisting of odd and even numbers of Dirac cones, respectively [4]. The existence of topological surface states is an important property of TIs, which is confirmed by nontrivial $Z_2$ invariants [4,5,9]. For systems with inversion symmetry, the $Z_2$ invariants can be calculated by the Fu-Kane parity criterion [4,10] based on the parity eigenvalues of the occupied bands at time-reversal invariant momenta (TRIM) points.

Band inversions play important roles in identifying various topological phases. Many strong topological nontrivial phases only have one band inversion, which is usually derived from different atomic orbital components; for example, $Bi_2Se_3$ [11] has a band inversion between the Bi $p$ and Se $p$ orbitals. In addition, one band inversion within a single $p$-orbital manifold has been demonstrated in transition-metal dichalcogenides [12,13] rather than being derived from different atomic orbitals. The transition-metal chalcogenide $Ta_2NiSe_5$ is a quasi-one-dimensional (quasi-1D) system. $Ta_2NiSe_5$ undergoes a structural phase transition from a high-temperature orthorhombic to a low-temperature monoclinic phase [14,15] at 328 K [16]. Interestingly, $Ta_2NiSe_5$ has been proposed as a candidate excitonic insulator at room temperature and ambient pressure [14,15,17-20]. The excitonic state at low temperature that is characterized by the flat top of the valence band was observed in angle-resolved photoemission spectroscopy experiments [17-19]. In recent years, two-dimensional topological excitonic insulators have attracted increasing interest [21,22]. However, the topological properties of $Ta_2NiSe_5$ have not yet been revealed. It would be interesting to check if $Ta_2NiSe_5$ could be a 3D topological excitonic insulator.

In this paper, we calculated the electronic structures of the orthorhombic and monoclinic phases of $Ta_2NiSe_5$ based on density functional theory (DFT). We find that a continuous direct spin-orbit coupling (SOC) gap exists in the whole Brillouin zone (BZ; even although they are not insulators). Therefore, the $Z_2$ invariants are well defined for the occupied bands below the gap in the orthorhombic and monoclinic phases of $Ta_2NiSe_5$. The calculated $Z_2$ invariants are (1; 000) in both phases, which indicates that they are STIs. We also calculated the topological surface states of the low-temperature monoclinic phase. Distinct from the previously proposed TIs with a single band inversion, multiple $pd$- and $dd$-type band inversions already occur at the Γ point even without SOC in both the orthorhombic and monoclinic phases of $Ta_2NiSe_5$. In addition, we find that an SOC-induced band inversion occurs at the Γ point in the monoclinic phase, whereas SOC does not induce a band inversion at the Γ point in the orthorhombic phase.

## II. COMPUTATIONAL METHOD

The first-principles calculations were performed in the WIEN2K [23] simulation package based on the full-potential linear augmented plane wave method. The Perdew-Burke-Ernzerhof parameterization of the generalized gradient approximation was used as the DFT exchange-correlation potential [24]. In the self-consistent calculations, the k-point sampling grid in the BZ was 28×28×6. The plane-wave cutoff was $K$max = $7.0/R_{MT}$. The energy convergence criterion was $10^{-6}$ Ry, and the charge convergence criterion was $10^{-4}$ e. The irreducible representations of electronic states were obtained with WIEN2K. Tight-binding Hamiltonians [25] were constructed using maximally localized Wannier functions [26-28] implemented in the WANNIER90 code [29] that is based on first-principles calculations. The surface Green's functions of the semi-infinite systems were obtained through an iterative method [30,31] that was implemented in the WANNIERTOOLS package [32].

## III. RESULTS AND DISCUSSION

$Ta_2NiSe_5$ crystallizes in a layered structure stacked by weak van der Waals interactions, and in each layer, single chain of Ni and double chains of Ta run along the $c$-axis of the lattice to form a quasi-1D chain structure [33], as shown in Fig. 1(a). The primitive cell is composed of four Ta ions, two Ni ions, and ten Se ions. All the Ta (Ni) ions are equivalent while there are three inequivalent sites of Se ions. $Ta_2NiSe_5$ undergoes a structural phase transition from a high-temperature orthorhombic phase to a low-temperature monoclinic phase [14,15] at 328 K [16]. The low-temperature monoclinic phase is synthesized in the centrosymmetric space group $C2/c$, and the high-temperature phase crystallizes in the orthorhombic phase with the centrosymmetric space group $Cmcm$. At room temperature, $Ta_2NiSe_5$ is slightly distorted from the orthorhombic phase with $β = 90°$ to a monoclinic phase with $β = 90.53°$, and they both have $α = γ = 90°$. We use the experimental crystal structures of the high-temperature orthorhombic [34] and low-temperature monoclinic [33] phases of $Ta_2NiSe_5$. The experimental lattice constants and atomic coordinates of the orthorhombic and monoclinic phases are listed in Tables I and II, respectively. The first BZ of $Ta_2NiSe_5$ is shown in Fig. 1(b). The $Z_2$ invariants can be determined by the parity products of the occupied bands at eight TRIM points in the centrosymmetric systems [4,10]. Among the eight TRIM points, there are two equivalent TRIM points denoted by $S$ and two equivalent TRIM points denoted by $R$; thus, the parity product of these four TRIM points is always 1. Furthermore, both the Z and T TRIM points are fourfold degenerate with the inversion matrix $\begin{pmatrix} 1 & 0 \\ 0 & -1 \end{pmatrix} \oplus \begin{pmatrix} 1 & 0 \\ 0 & -1 \end{pmatrix}$. Therefore, the parity product that includes the Z and T TRIM points is 1. Thus, only the parities of the occupied bands at the Γ and Y TRIM points make sense to determine the topologically nontrivial nature of $Ta_2NiSe_5$.

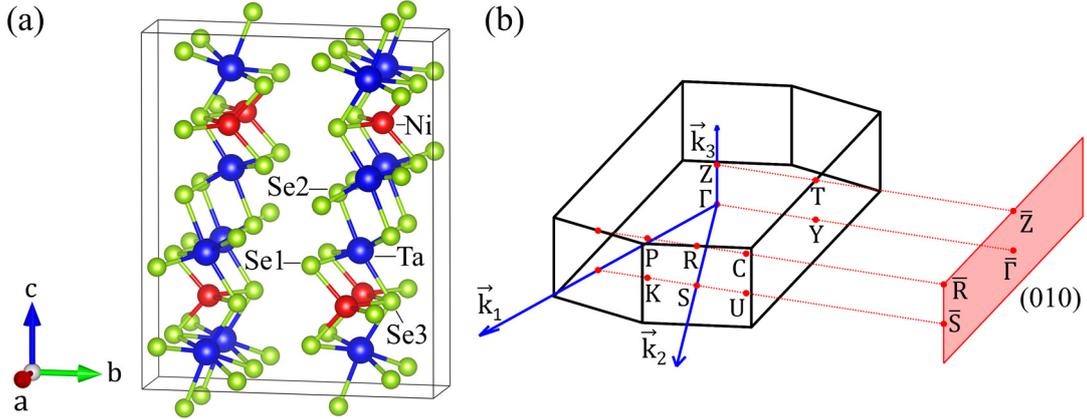

FIG. 1. (a) Side view of the crystal structure of the orthorhombic phase of Ta2NiSe5. The crystal structure of the monoclinic phase is visually indistinguishable from that of the orthorhombic phase, since the angle β between the lattice vectors *a* and *c* is 90.53° in the monoclinic phase, while β = 90° in the orthorhombic phase. (b) Bulk Brillouin zone (BZ), projected surface BZ, and reciprocal points of the orthorhombic phase of Ta2NiSe5. Correspondingly, the bulk BZ of the monoclinic phase is almost indistinguishable from that of the orthorhombic phase. Γ (0, 0, 0), K (0.25, 0.25, 0), S (0, 0.5, 0), R (0, 0.5, 0.5), P (0.25, 0.25, 0.5), Z (0, 0, 0.5), Y (-0.5, 0.5, 0), U (-0.25, 0.75, 0), C (-0.25, 0.75, 0.5), T (-0.5, 0.5, 0.5). The reciprocal points in the BZ are given

as linear combinations of the reciprocal lattice vectors of the (based-centered) primitive cell.

TABLE I. Experimental crystallographic data of the orthorhombic [34] and monoclinic [33] phases of $Ta_2NiSe_5$.

| Phase | Orthorhombic | Monoclinic |
|---|---|---|
| Space group | *Cmcm* | *C2/c* |
| $a$ (Å) | 3.5029 | 3.496 |
| $b$ (Å) | 12.8699 | 12.829 |
| $c$ (Å) | 15.6768 | 15.641 |
| $\beta$ (°) | 90 | 90.53 |
| $V$ (Å$^3$) | 706.74 | 701.42 |

TABLE II. Experimental atomic coordinates for the orthorhombic [34] and monoclinic phases [33] of $Ta_2NiSe_5$.

| | | Orthorhombic | | | Monoclinic | | |
|---|---|---|---|---|---|---|---|
| Atom | Site | $x$ | $y$ | $z$ | $x$ | $y$ | $z$ |
| Ni | *4c* | 0 | 0.70096 | 0.25 | 0 | 0.70113 | 0.25 |
| Ta | *8f* | 0 | 0.221158 | 0.110222 | 0.99207 | 0.221349 | 0.110442 |
| Se(1) | *8f* | 0 | 0.580461 | 0.137726 | 0.0053 | 0.580385 | 0.137979 |
| Se(2) | *8f* | 0 | 0.14583 | 0.950662 | 0.99487 | 0.145648 | 0.950866 |
| Se(3) | *4c* | 0 | 0.32679 | 0.25 | 0 | 0.32714 | 0.25 |

The band structures of the orthorhombic and monoclinic phases of $Ta_2NiSe_5$ along the high-symmetry paths in the BZ are shown in Fig. 2. There are three band crossings near the Fermi level along the $Z$-$\Gamma$, $\Gamma$-$Y$ and $T$-$Y$ lines in the band structure of the orthorhombic phase when SOC is ignored [Fig. 2(a)]. When SOC is considered, a continuous direct gap between the conduction and valence bands opens in the entire BZ with a gap minimum located around the $\Gamma$ point as shown in Fig. 2(b). Here, *continuous direct gap* means that at every momentum **k** throughout the BZ is a direct gap between the conduction and valence bands, whereas the indirect gap is allowed to close. In contrast, in the monoclinic phase, the band structure has a continuous direct gap between the conduction and valence bands in the BZ without SOC [Fig. 2(c)] and with

SOC [Fig. 2(d)], and the gap minimum is also around the Γ point.

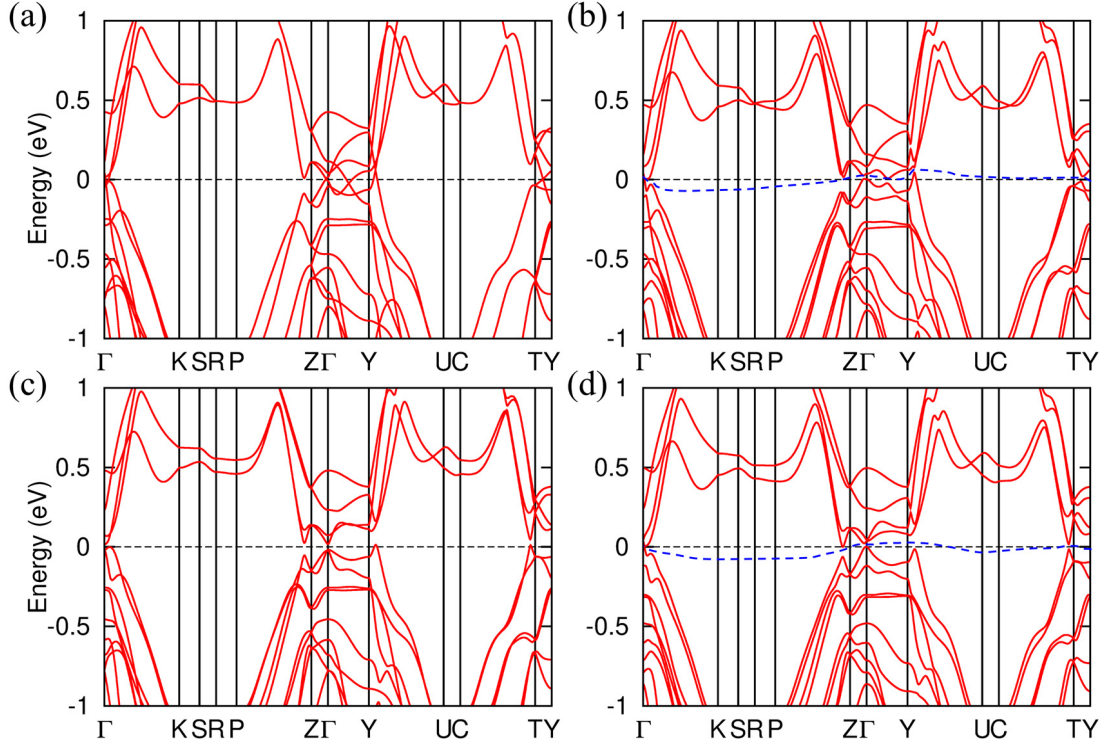

FIG. 2. Band structures of the (a) and (b) orthorhombic and (c) and (d) monoclinic phases of Ta$_2$NiSe$_5$ (a) and (c) without spin-orbit coupling (SOC) and (b) and (d) with SOC. The blue dashed lines in (b) and (d) show the continuous direct gap between the conduction and valence bands in the entire Brillouin zone (BZ) and represent the assumed Fermi level across the gap. The dashed straight line indicates the actual Fermi level.

A direct way to determine whether a system with TRS has nontrivial topological properties is to calculate the $Z_2$ invariants. There are four $Z_2$ invariants denoted by ($v_0$; $v_1$, $v_2$, $v_3$) that distinguish the ordinary insulators from WTIs and STIs in three dimensions. STIs are indicated by the topological invariant $v_0$ = 1. For systems with inversion symmetry, the $Z_2$ invariants can be determined by the Fu-Kane parity criterion [4,10], which is based on the parity eigenvalues of the occupied Bloch wave functions at the TRIM points. When SOC is included, the band structures of the orthorhombic and monoclinic phases of Ta$_2$NiSe$_5$ have a continuous direct gap between the conduction and valence bands in t ⟨he entire BZ. Therefore, the Fu-Kane parity criterion can be defined for the parities of the occupied bands below the gap. In the band structures of the orthorhombic and monoclinic phases of $Ta_2NiSe_5$, there is a large continuous gap in the entire BZ < -7 eV (relative to the Fermi level), which suggests that no band inversions occur between the bands below and above the large gap. Therefore, the parity product of the bands below the large gap at eight TRIM points is 1. The number of bands between -7 eV and the assumed Fermi level with odd/even parity at the TRIM points, without (w/o) SOC and with (w/) SOC, respectively, are presented in Table III. The $Z_2$ invariants of the orthorhombic and monoclinic phases of $Ta_2NiSe_5$ are both calculated to be (1; 000), which indicates that they are both STIs.

TABLE III. Number of bands between -7 eV and the assumed Fermi level with odd/even parity at the TRIM points, without (w/o) SOC and with (w/) SOC in the orthorhombic and monoclinic phases of $Ta_2NiSe_5$. Among the eight TRIM points, (0.5, 0, 0) and (0, 0.5, 0) are equivalent TRIM points while (0.5, 0, 0.5) and (0, 0.5, 0.5) are equivalent TRIM points. Therefore we only show the results for S (0, 0.5, 0) and R (0, 0.5, 0.5).

| $Ta_2NiSe_5$ | TRIM points | Γ (0,0,0) | S (0,0.5,0) | Z (0,0,0.5) | Y (0.5,0.5,0) | R (0,0.5,0.5) | T (0.5,0.5,0.5) |
|---|---|---|---|---|---|---|---|
| Orthorhombic | w/o SOC | 19/21 | 20/20 | 20/20 | 20/20 | 20/20 | 20/20 |
| | w/SOC | 19/21 | 20/20 | 20/20 | 20/20 | 20/20 | 20/20 |
| Monoclinic | w/o SOC | 20/20 | 20/20 | 20/20 | 20/20 | 20/20 | 20/20 |
| | w/SOC | 19/21 | 20/20 | 20/20 | 20/20 | 20/20 | 20/20 |

To understand the origin of this nontrivial band topology, i.e., nontrivial $Z_2$, it is important to elucidate the mechanism of the band inversions at the Γ and Y TRIM points. To this end, we analyzed eight important bands around the Fermi level which are denoted by numbers 1-8 according to their orbital character and irreducible representations, as shown in Figs. 3, 4, and S1-S6 in the Supplemental Material [35]. We notice that the distribution of these bands varies rapidly near $\Gamma$ and $Y$ points which are traced by analyzing the projected band structures (the so-called fat-bands, shown in Figs. S3-S6 in the Supplemental Material [35]) and irreducible representations of each band (shown in Figs. S1 and S2 in the Supplemental Material [35]) in the orthorhombic and monoclinic phases of Ta$_2$NiSe$_5$. We calculated the projected orbital weights on each nonequivalent Ni, Ta and Se atom. Bands 1, 3, 5, and 7 are all dominated by the Ta $d_{z^2}$ orbital, as shown in Fig. 3 for the orthorhombic phase and Fig. 4 for the monoclinic phase. Band 2 is mainly derived from Se(1) $p_z$ hybridizing with Se(2) $p_z$ orbitals, as shown in Figs. S3(b), S3(c), S4(b), and S4(c) in the Supplemental Material [35] for the orthorhombic phase, and Figs. S5(b), S5(c), S6(b), and S6(c) in the Supplemental Material [35] for the monoclinic phase. Band 4 is dominated by Se(2) $p_y$ hybridizing with Se(3) $p_x$ orbitals, as shown in Figs. S3(d), S3(e), S4(d), and S4(e) in the Supplemental Material [35] for the orthorhombic phase, and Figs. S5(d), S5(e), S6(d), and S6(e) in the Supplemental Material [35] for the monoclinic phase. Band 6 primarily originates from Ni $d_{yz}$ hybridizing with Se(3) $p_y$ orbitals, as shown in Figs. S3(f), S3(g), S4(f), and s4(g) in the Supplemental Material [35] for the orthorhombic phase, and Figs. S5(f), S5(g), S6(f), and S6(g) in the Supplemental Material [35] for the monoclinic phase. Band 8 is dominated by the Ni $d_{x^2y^2}$ orbital, as shown in Figs. S3(h) and S4(h) in the Supplemental Material [35] for the orthorhombic phase, and Figs. S5(h) and S6(h) in the Supplemental Material [35] for the monoclinic phase. The sizes of the red, blue and green curves represent the weights of the Ta $d$, Ni $d$ and Se $p$ orbitals, respectively, as shown in Figs. S3-S6 in the Supplemental Material [35].

We also label the parities of the bands at the $\Gamma$ and $Y$ points, as shown in Figs. 3 and 4. In addition, we label in Figs. S1 and S2 in the Supplemental Material [35] the eight (non-spin-orbital coupled) bands with irreducible representations.

Along the $K$-$\Gamma$ line, we analyze the band inversions at the $\Gamma$ point and note that four band inversions already occur even without SOC, which is different from the previously proposed TIs [11,36], which have only a single SOC-induced band inversion. The band inversion between bands 1 (3) and 2 (4) occurs above (below) the Fermi level, as shown in Fig. 3(a). The band inversions between bands 5 and 6 and between bands 7 and 8 occur on both sides of the Fermi level, as shown in Fig. 3(a). Therefore, *pd*-type [37] and *dd*-type band inversions both occur in the orthorhombic phase of Ta$_2$NiSe$_5$, while many topologically nontrivial phases exhibit *sp*- or *pp*-type band inversion [11,38-40].

At the $\Gamma$ point, bands 2, 6, and 8 with the irreducible representations $\Gamma_1^-$, $\Gamma_4^-$, and $\Gamma_2^-$, respectively, all cross the Fermi level from below along the *K*-$\Gamma$ line, as shown in Figs. 3(a) and S1 in the Supplemental Material [35]. On the other hand, bands 3, 5, and 7 with the irreducible representations $\Gamma_3^-$, $\Gamma_1^+$, and $\Gamma_2^-$, respectively, all cross the Fermi level from above along the *K*-$\Gamma$ line. Therefore, the number of odd parities moving above the Fermi level, i.e. three, is not equal to the number of odd parities sinking below the Fermi level, i.e., two, at the $\Gamma$ point. This explains the overall odd parity of the occupied bands at the $\Gamma$ point shown in Table III.

Along the *U*-*Y* line, the band inversion between bands 5 and 6 at the *Y* point occurs above the Fermi level. At the *Y* point, band 3 with the irreducible representation $\Gamma_4^+$ crosses the Fermi level from above, and band 6 with the irreducible representation $\Gamma_3^+$ crosses the Fermi level from below along the *U*-*Y* line, as shown in Figs. 3(a) and S1 in the Supplemental Material [35]. Therefore, the number of odd parities of the occupied states at the *Y* point remains unchanged.

The inclusion of SOC in the orthorhombic phase of Ta$_2$NiSe$_5$ opens a continuous direct gap between the conduction and valence bands in the entire BZ but does not change the energy order of the bands at the TRIM points, as shown in Fig. 3(b). In addition, all the irreducible representations of these bands without and with SOC at the $\Gamma$ and *Y* points are presented in Table IV.

Similar analyses at the $\Gamma$ and *Y* points are carried out for the monoclinic phase of Ta$_2$NiSe$_5$. There are still four band inversions at the $\Gamma$ point along the *K*-$\Gamma$ line when SOC is ignored. Like the orthorhombic phase at the $\Gamma$ point, the band inversion between bands 1 and 2 occurs above the Fermi level, and the band inversion between bands 3 and 4 occurs below the Fermi level, as shown in Fig. 4(a).

Different from the orthorhombic phase at the $\Gamma$ point, band 5 is above the Fermi level,

and band 8 is below the Fermi level. Therefore, at the Γ point, the band inversion between bands 5 and 6 occurs above the Fermi level, and the band inversion between bands 7 and 8 occurs below the Fermi level, as shown in Fig. 4(a).

At the Γ point, band 2 (6) with the irreducible representation $\Gamma_1^-$ ($\Gamma_2^-$) crosses the Fermi level from below along the K-Γ line, while band 3 (7) with the irreducible representation $\Gamma_2^-$ ($\Gamma_1^-$) crosses the Fermi level from above along the K-Γ line, as shown in Figs. 4(a) and S2 in the Supplemental Material [35]. The number of odd parities moving above the Fermi level is equal to the number of odd parities sinking below the Fermi level; thus, the number of odd parities of the occupied states at the Γ point is unchanged.

Along the U-Y line, the band inversion between bands 5 and 6 at the Y point also occurs above the Fermi level, which is analogous to that of the orthorhombic phase. At the Y point, band 3 (6) with the irreducible representation $\Gamma_2^+$ crosses the Fermi level from above (below) along the U-Y line, as shown in Figs. 4(a) and S2 in the Supplemental Material [35]. Therefore, the number of odd parities of the occupied states at the Y point stays the same. The invariant number of odd parities of the occupied states at the Γ and Y points leads to the trivial $Z_2$. When SOC is taken into consideration, we find an inversion between band 5 with even parity and band 8 with odd parity that occurs at the Γ point by comparing Fig. 4(a) and (b). Therefore, when SOC is included, the parities of the occupied states at Γ lead to a topologically nontrivial $Z_2$ in the monoclinic phase of Ta$_2$NiSe$_5$, which is like the orthorhombic phase.

All the irreducible representations of these bands without and with SOC at the Γ and Y points are also presented in Table IV. Notably, Table III shows that, when SOC is included, the number of occupied bands > -7 eV with odd/even parity only changes at the Γ point in the monoclinic phase of Ta$_2$NiSe$_5$; this behavior is attributed to the SOC-induced band inversion at the Γ point, like that of Bi$_2$Se$_3$ [11,36]. The above result agrees well with our band inversion and parity analysis. The strong topological invariant $v_0$ also changes from 0 to 1, which implies that the monoclinic phase of Ta$_2$NiSe$_5$ is trivial when SOC is ignored but is nontrivial when SOC is considered.

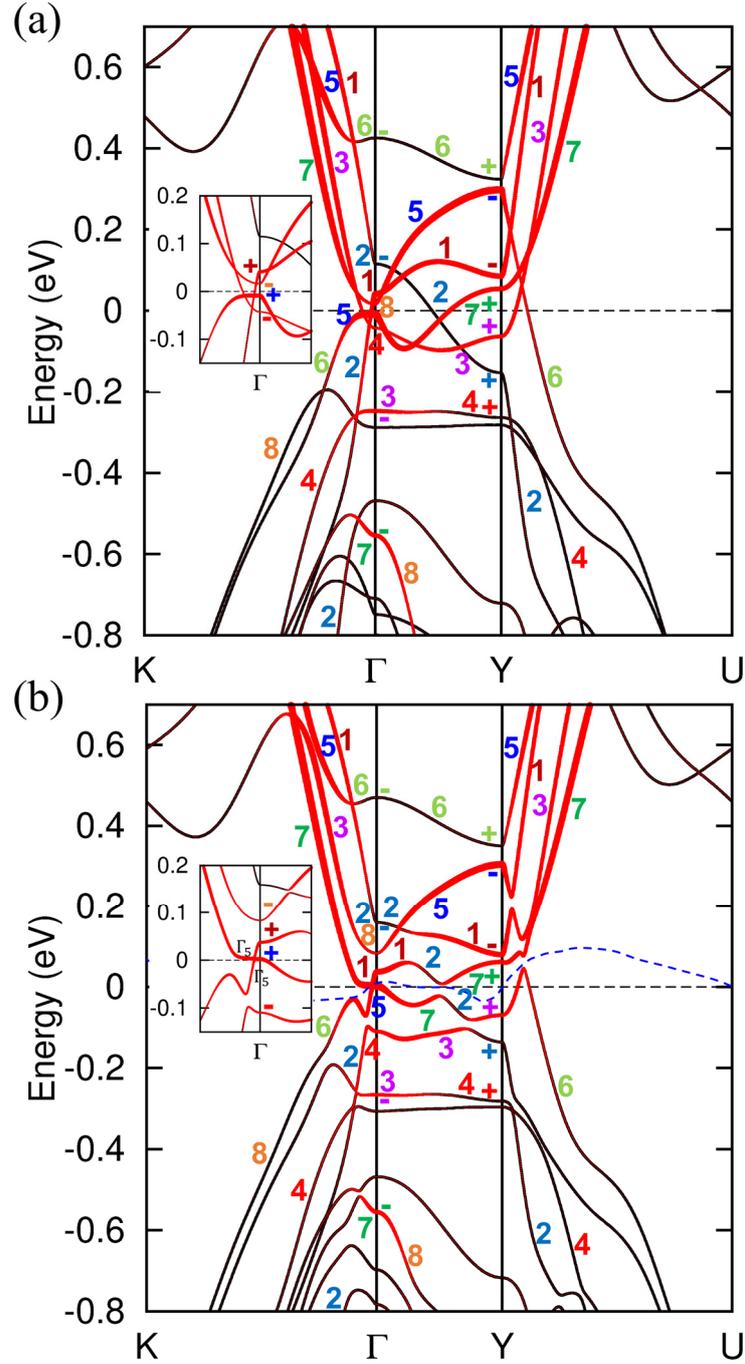

FIG. 3. Band structures of the orthorhombic phase of Ta$_2$NiSe$_5$ along $K$-$\Gamma$-$Y$-$U$ $k$-path (a) without spin-orbit coupling (SOC) and (b) with SOC. The size of the red curves represents the weight of the Ta $d_{z^2}$ orbital. The numbers 1-8 denote the eight bands. The parities are labeled for the bands without and with SOC. The blue dashed line in (b) represents the assumed Fermi level across the gap. The dashed straight line indicates the actual Fermi level.

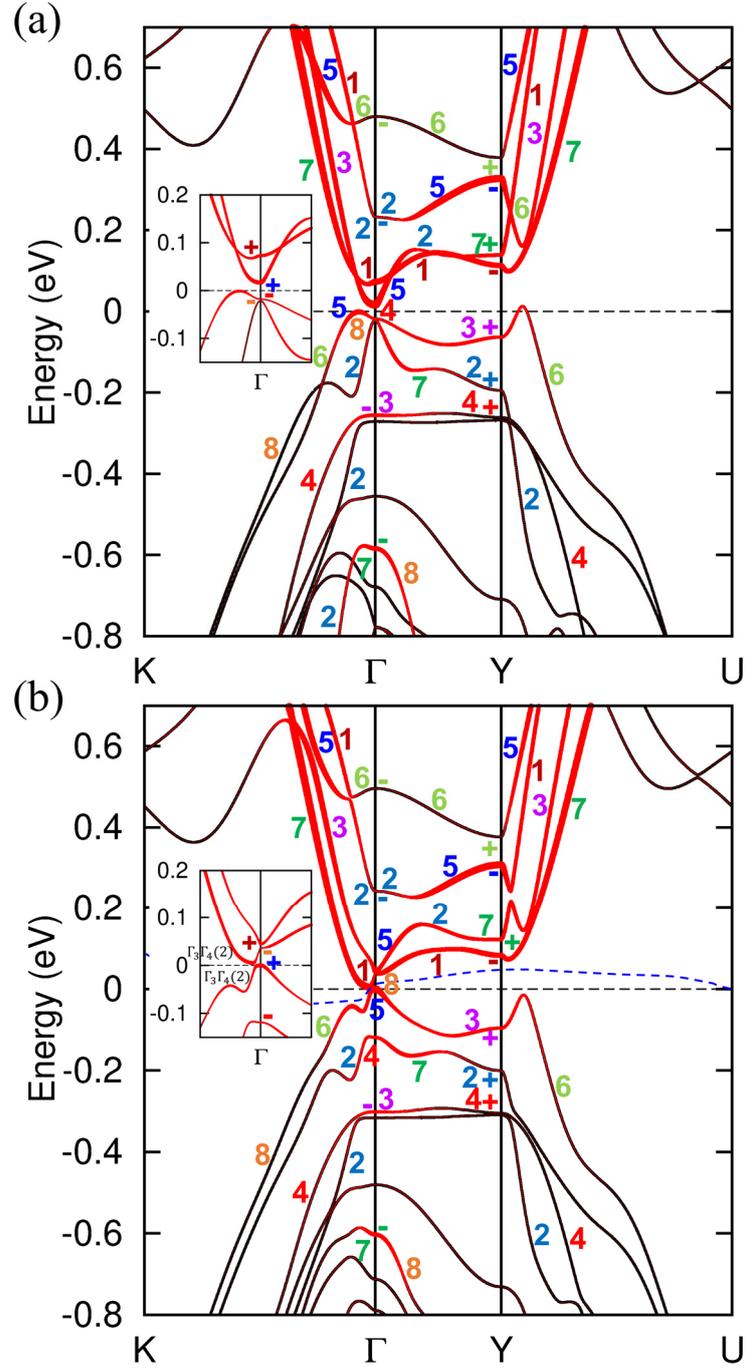

FIG. 4. Band structures of the monoclinic phase of Ta$_2$NiSe$_5$ along $K$-$\Gamma$-$Y$-$U$ k-path (a) without spin-orbit coupling (SOC) and (b) with SOC. The size of the red curves represents the weight of the Ta $d_{z^2}$ orbital. The numbers 1-8 denote the eight bands. The parities are labeled for the bands without and with SOC. The blue dashed line in (b) represents the assumed Fermi level across the gap. The dashed straight line indicates the actual Fermi level.

TABLE IV. Irreducible representations at the Γ and Y points of the bands crossing the Fermi level or Fermi curve. Up (down) arrow indicates the band moves above (below) the Fermi level/curve at the Γ (Y) point along $K$-Γ ($U$-Y) direction.

| Orthorhombic phase of Ta$_2$NiSe$_5$ | | | | Monoclinic phase of Ta$_2$NiSe$_5$ | | | |
|---|---|---|---|---|---|---|---|
| w/o SOC | | w/ SOC | | w/o SOC | | w/ SOC | |
| $K$-Γ | Γ | $K$-Γ | Γ | $K$-Γ | Γ | $K$-Γ | Γ |
| #2↑ | $\Gamma_1^-$ | #2↑ | $\Gamma_5^-$ | #2↑ | $\Gamma_1^-$ | #2↑ | $\Gamma_3^- \oplus \Gamma_4^-$ |
| #6↑ | $\Gamma_4^-$ | #6↑ | $\Gamma_5^-$ | #6↑ | $\Gamma_2^-$ | #6↑ | $\Gamma_3^- \oplus \Gamma_4^-$ |
| #8↑ | $\Gamma_2^-$ | #8↑ | $\Gamma_5^-$ | | | #8↑ | $\Gamma_3^- \oplus \Gamma_4^-$ |
| Fermi level | | Fermi curve | | Fermi level | | Fermi curve | |
| #3↓ | $\Gamma_3^-$ | #3↓ | $\Gamma_5^-$ | #3↓ | $\Gamma_2^-$ | #3↓ | $\Gamma_3^- \oplus \Gamma_4^-$ |
| #5↓ | $\Gamma_1^+$ | #5↓ | $\Gamma_5^+$ | | | #5↓ | $\Gamma_3^+ \oplus \Gamma_4^+$ |
| #7↓ | $\Gamma_2^-$ | #7↓ | $\Gamma_5^-$ | #7↓ | $\Gamma_1^-$ | #7↓ | $\Gamma_3^- \oplus \Gamma_4^-$ |
| | | | | | | | |
| $U$-Y | Y | $U$-Y | Y | $U$-Y | Y | $U$-Y | Y |
| #6↑ | $\Gamma_3^+$ | #6↑ | $\Gamma_5^+$ | #6↑ | $\Gamma_2^+$ | #6↑ | $\Gamma_3^+ \oplus \Gamma_4^+$ |
| Fermi level | | Fermi curve | | Fermi level | | Fermi curve | |
| #3↓ | $\Gamma_4^+$ | #3↓ | $\Gamma_5^+$ | #3↓ | $\Gamma_2^+$ | #3↓ | $\Gamma_3^+ \oplus \Gamma_4^+$ |

An important property of topological nontrivial phases is the existence of topological surface states. Because the surface states of the low-temperature monoclinic phase of Ta$_2$NiSe$_5$ are easier to be observed in experiments, we calculated the surface states of the monoclinic phase of Ta$_2$NiSe$_5$ on the (010) surface (Fig. 5). We can clearly see that the Dirac cones occur at the $\bar{\Gamma}$ point on the (010) surface. Thus, the surface-state calculations are consistent with the odd $Z_2$ invariant and further confirm that the monoclinic phase of Ta$_2$NiSe$_5$ is a strong TI. In addition, the monoclinic phase of Ta$_2$NiSe$_5$ has been demonstrated to be an excitonic insulator at room temperature and ambient pressure [14,15,17-20]. Therefore, the monoclinic phase of Ta$_2$NiSe$_5$ has promising potential to become a 3D topological excitonic insulator.

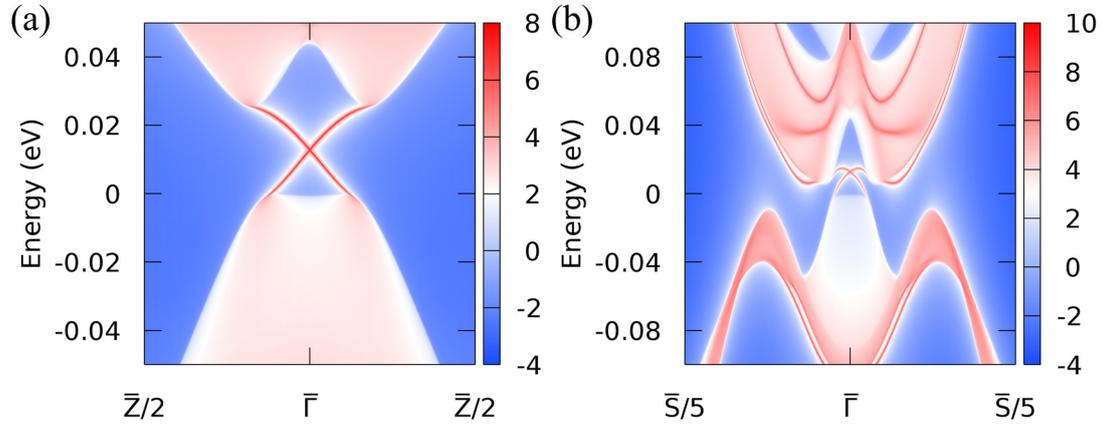

FIG. 5. Topological surface states of the monoclinic phase of Ta$_2$NiSe$_5$ on the (010) surface. The time-reversal invariant momenta (TRIM) points on the two-dimensional (2D) Brillouin zone (BZ) of the projected surface are $\bar{\Gamma}$ (0, 0), $\bar{Z}$ (0, 0.5), and $\bar{S}$ (0.5, 0).

IV. CONCLUSIONS

In summary, the topologically nontrivial electronic structures of the orthorhombic and monoclinic phases of $Ta_2NiSe_5$ were obtained based on first-principles calculations. These two phases of $Ta_2NiSe_5$ are STIs supported by $Z_2$ invariants (1; 000). The topological surface states of the low-temperature monoclinic phase were also calculated. A distinct property is that there are four band inversions at the Γ point in the orthorhombic and monoclinic phases of $Ta_2NiSe_5$, leading to a reduction by one of the number of occupied bands with odd parity at the Γ point which gives rise to the topologically nontrivial $Z_2$ invariant. The multiple *pd*- and *dd*-type band inversions at the Γ point in $Ta_2NiSe_5$ are different from the previously proposed TI phases with only a single *sp*- or *pp*-type band inversion. In addition, the mechanisms of the nontrivial band topology of the orthorhombic and monoclinic phases of $Ta_2NiSe_5$ are different. SOC does not induce a band inversion at the Γ point in the orthorhombic phase, while a band inversion at the Γ point is induced by SOC in the monoclinic phase. More importantly, $Ta_2NiSe_5$ has been previously reported as an excitonic insulator. Combined with our results which demonstrate that $Ta_2NiSe_5$ is a STI, $Ta_2NiSe_5$ is expected to be a promising 3D topological excitonic insulator. Given the approximate nature of the exchange-correlation functional used in the DFT calculations, our results provide a preliminary reference for experimentally and theoretically exploring the topological property of $Ta_2NiSe_5$ in the future.

Acknowledgements: This paper was supported by the National Natural Science Foundation of China (Grants No. 12074041 and No. 11674030), the Fundamental Research Funds for the Central Universities (Grant No. 310421113), the National Key Research and Development Program of China through Contract No. 2016YFA0302300, and the start-up funding of Beijing Normal University.

#X. B. M. and G. W. W. contributed equally to this work. *Requests for materials should be addressed to Z. P. Y. at yinzhiping@bnu.edu.cn.

# Supplemental Material for
# Ta$_2$NiSe$_5$: a candidate topological excitonic insulator with multiple band inversions

Xiaobo Ma[#], Guangwei Wang[#], Huican Mao, Zhihong Yuan, Tianye Yu, Rui Liu, Yiran Peng, Pengyu Zheng, and Zhiping Yin*

*Department of Physics and Center for Advanced Quantum Studies, Beijing Normal University, Beijing 100875, China*

In this Supplemental Material, we present eight important bands around the Fermi level which are denoted by numbers 1-8 according to projected band structures (the so-called fat-bands, shown in Figs. S3-S6) and irreducible representations of each band without SOC (shown in Figs. S1 and S2) in the orthorhombic and monoclinic phases of Ta$_2$NiSe$_5$.

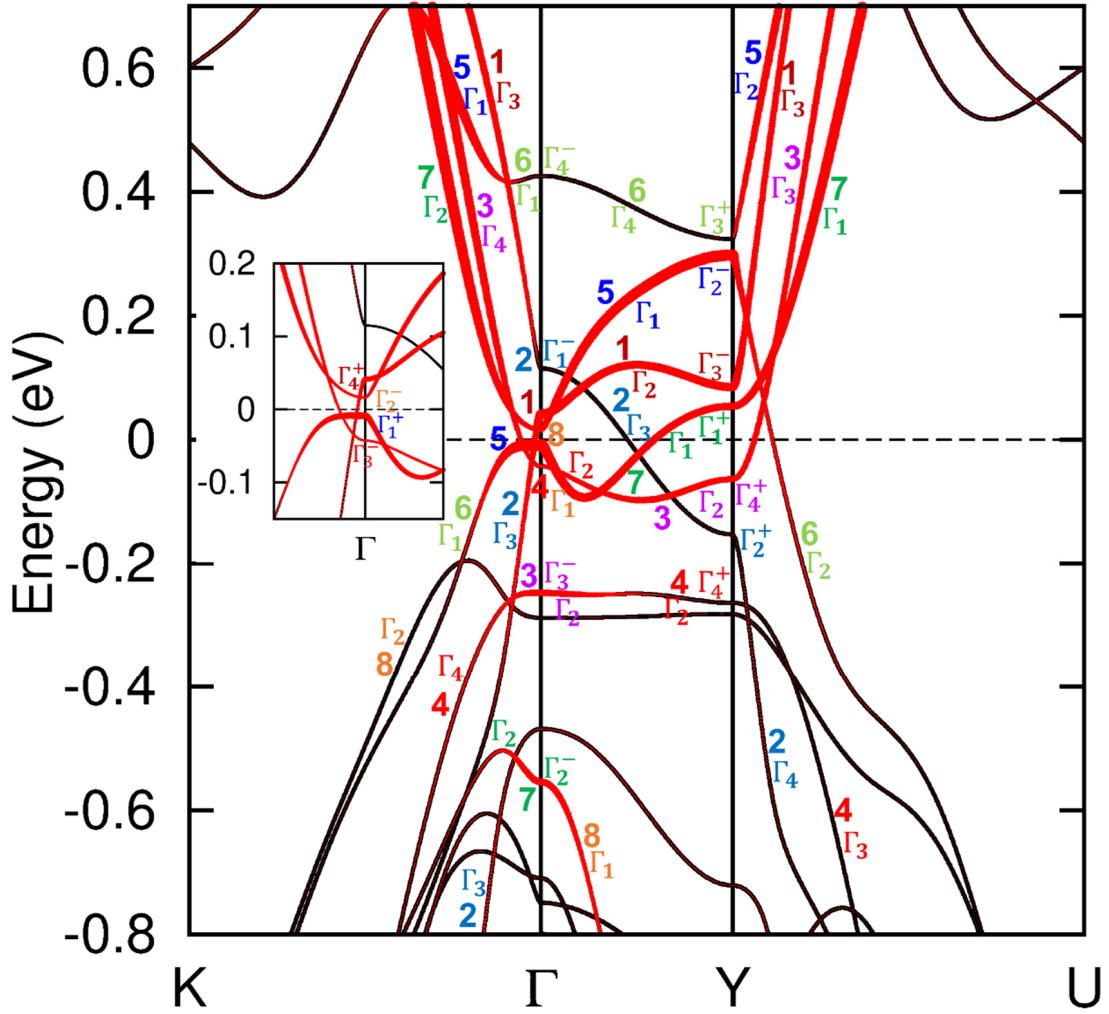

FIG. S1. Band structure of the orthorhombic phase of Ta$_2$NiSe$_5$ along *K*-Γ-*Y*-*U* *k*-path without SOC. The size of the red curves represents the weight of the Ta $d_{z^2}$ orbital. The numbers 1-8 denote the eight bands. The irreducible representations are labeled for the bands without SOC.

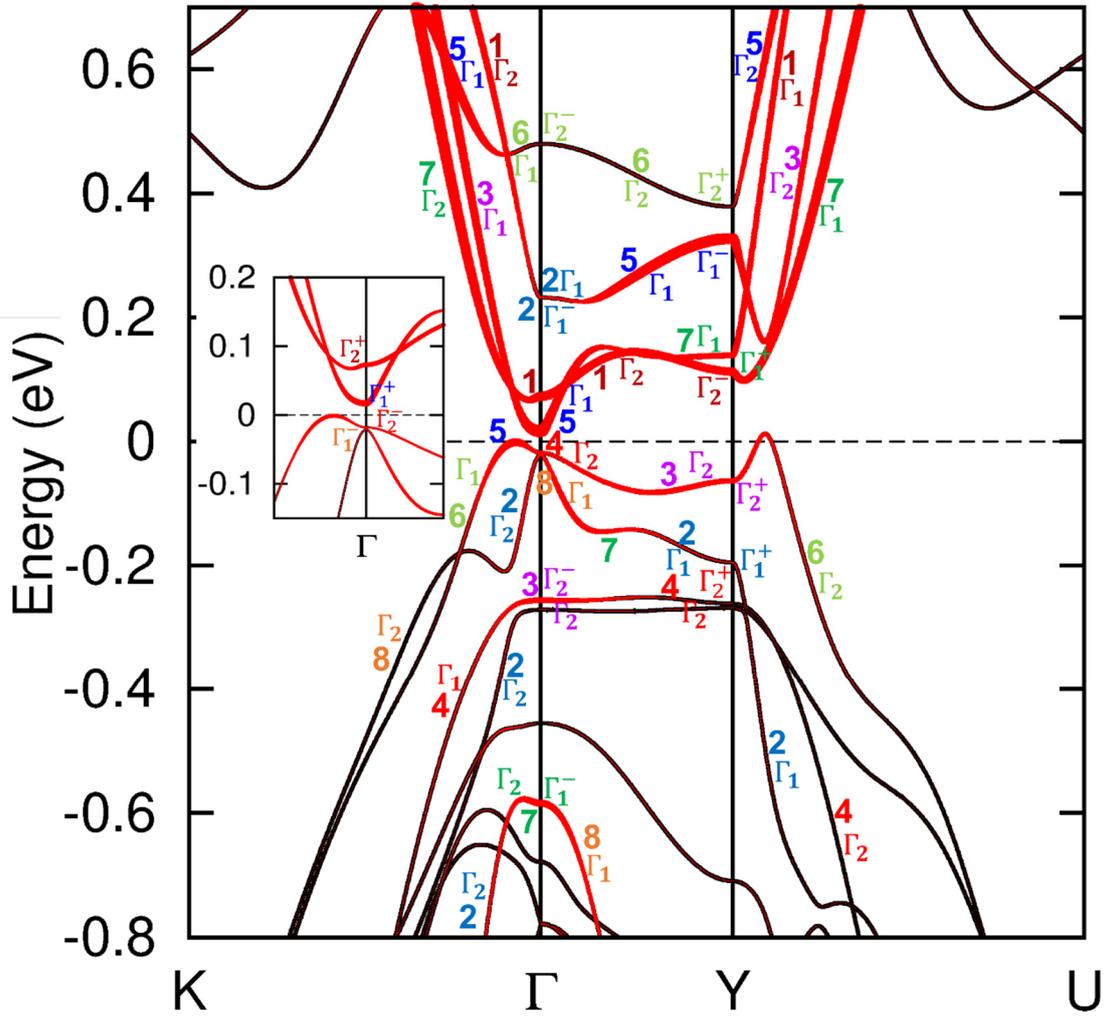

FIG. S2. Band structure of the monoclinic phase of Ta$_2$NiSe$_5$ along $K$-$\Gamma$-$Y$-$U$ $k$-path without SOC. The size of the red curves represents the weight of the Ta $d_{z^2}$ orbital. The numbers 1-8 denote the eight bands. The irreducible representations are labeled for the bands without SOC.

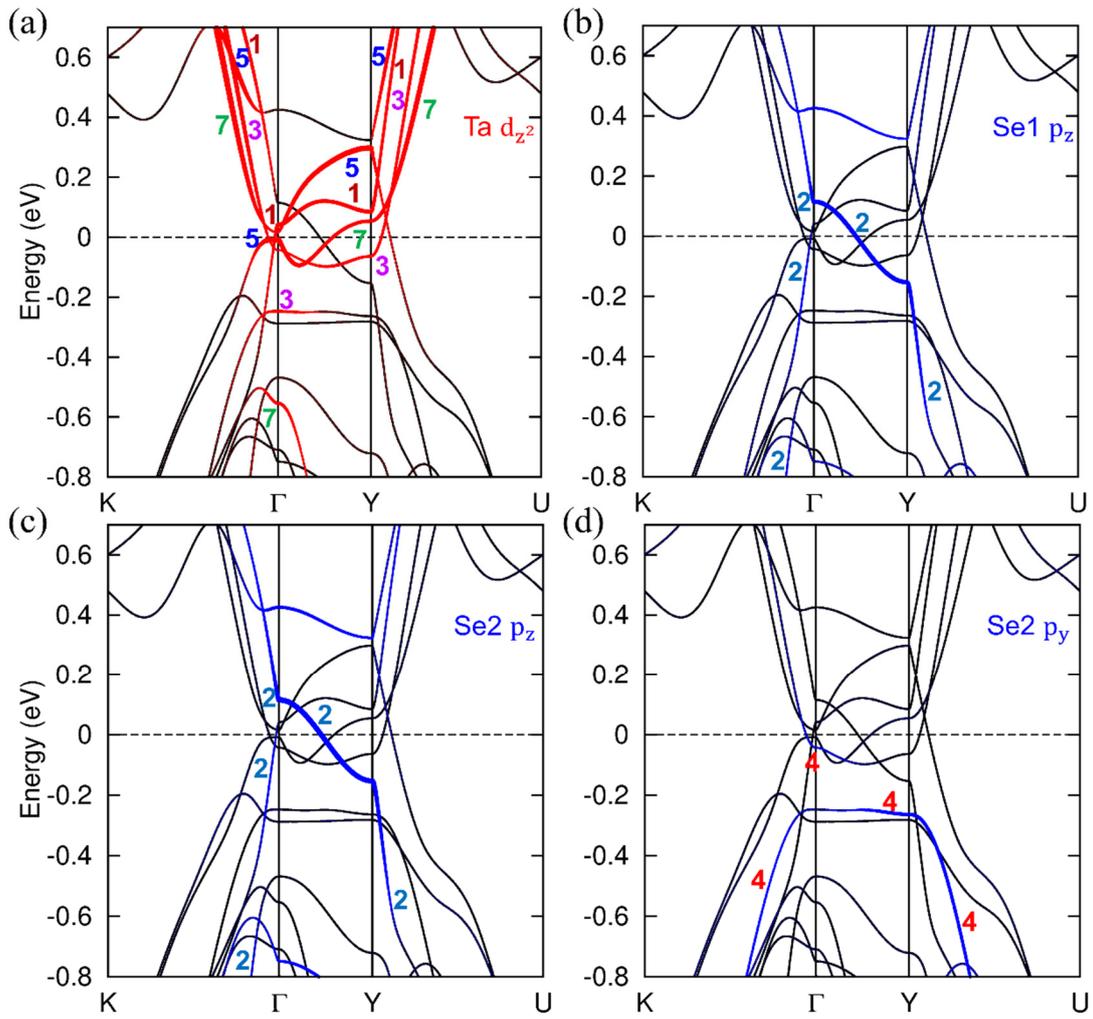

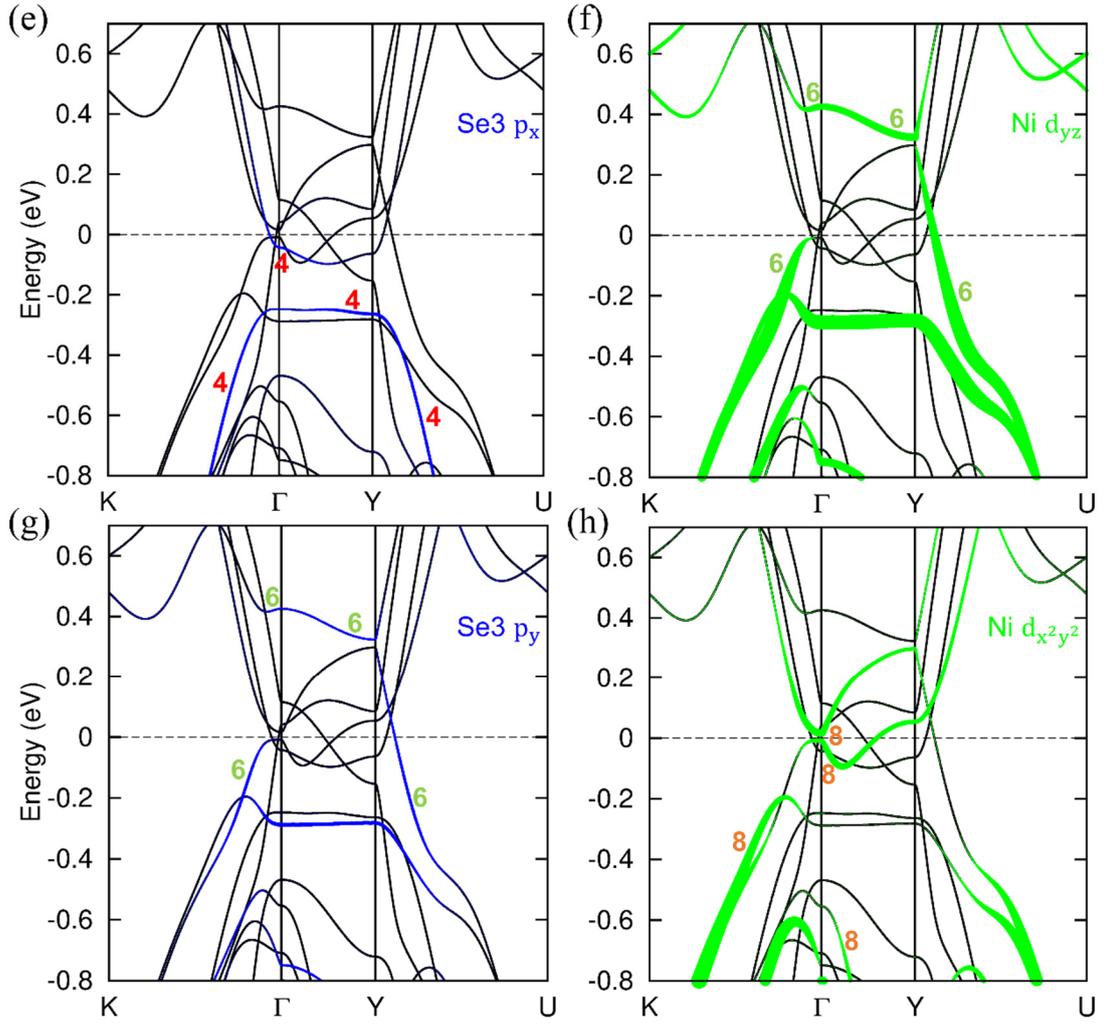

FIG. S3. Projected band structures of the orthorhombic phase of $Ta_2NiSe_5$ without SOC (a)-(h). The sizes of the red, blue and green curves represent the weights of the Ta $d$, Ni $d$ and Se $p$ orbitals, respectively. The numbers 1-8 denote the eight bands.

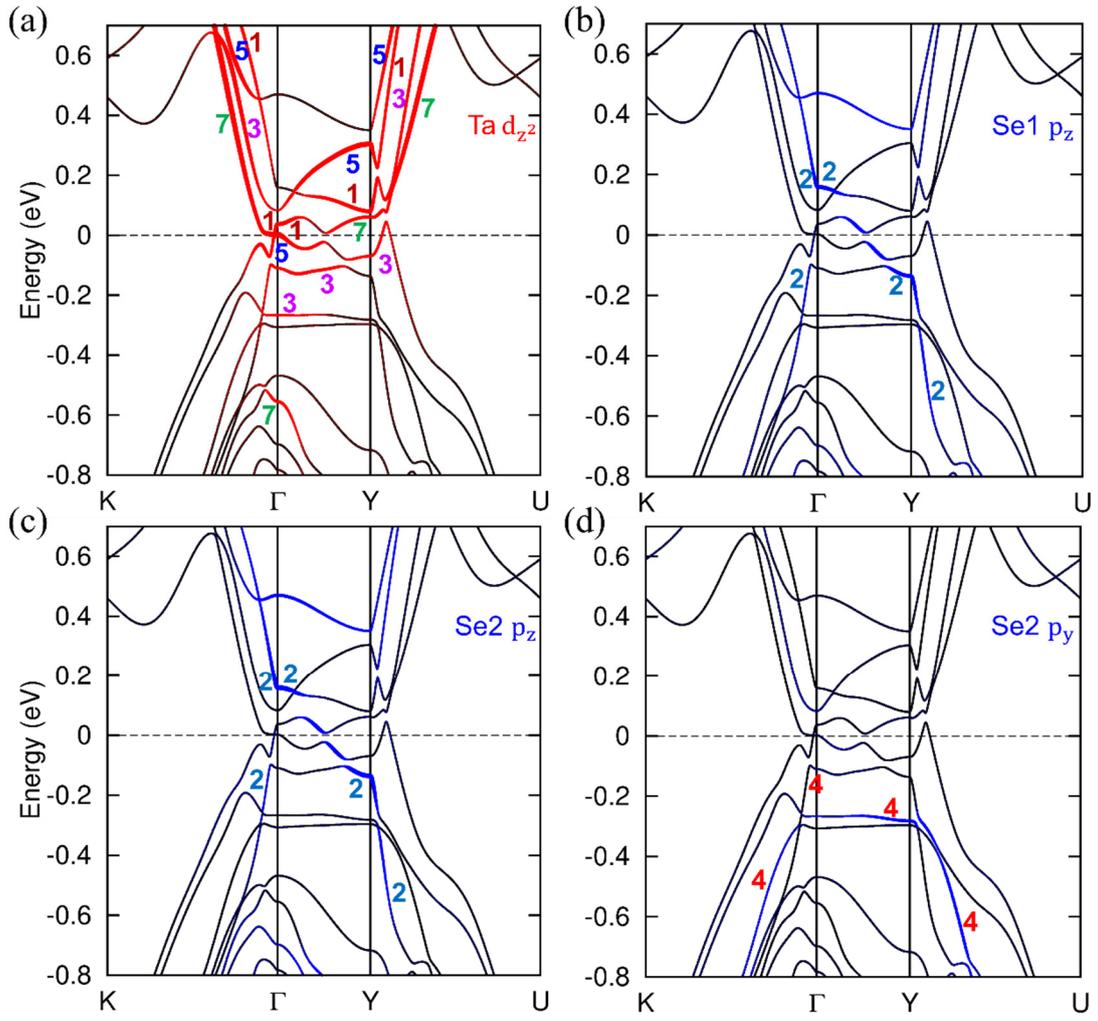

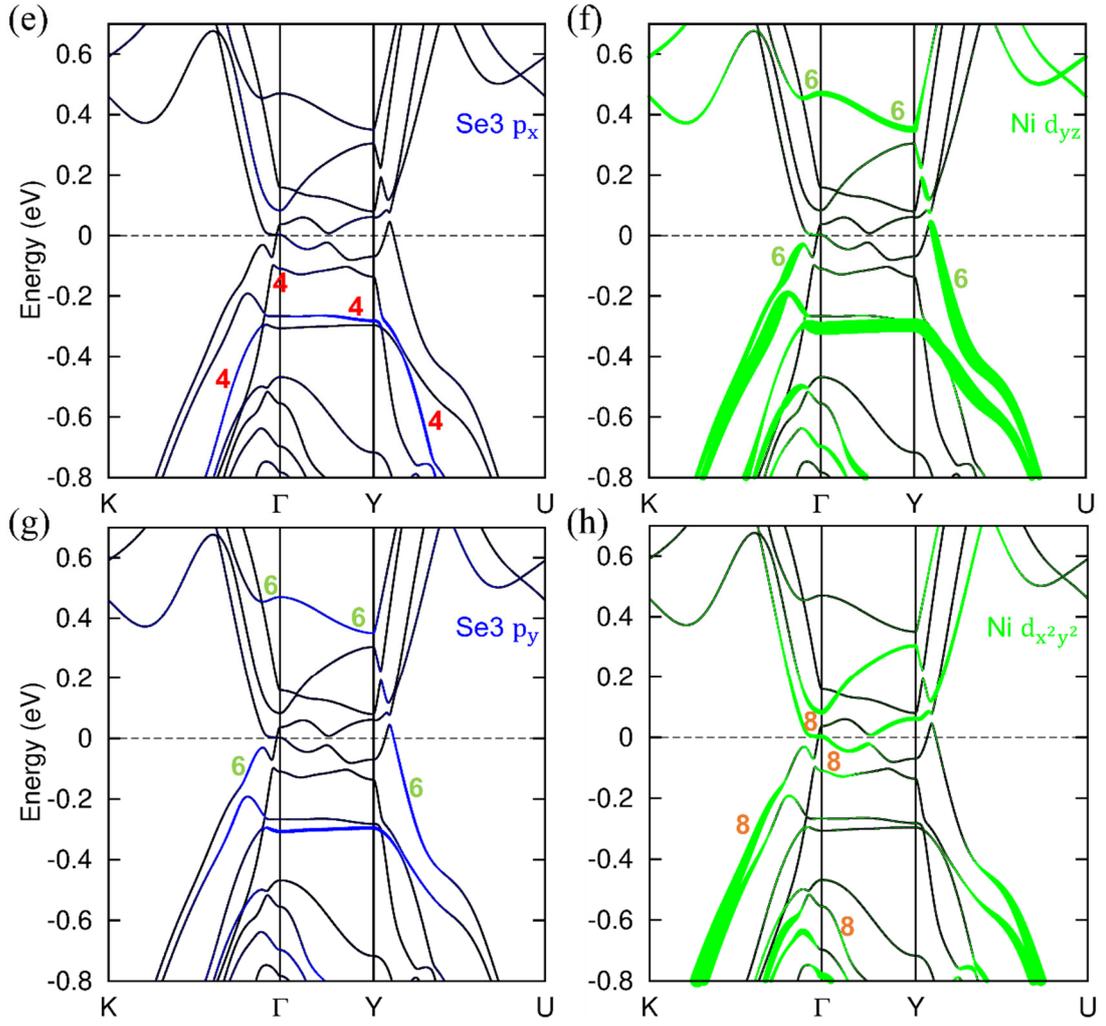

FIG. S4. Projected band structures of the orthorhombic phase of Ta$_2$NiSe$_5$ with SOC (a)-(h). The sizes of the red, blue and green curves represent the weights of the Ta $d$, Ni $d$ and Se $p$ orbitals, respectively. The numbers 1-8 denote the eight bands.

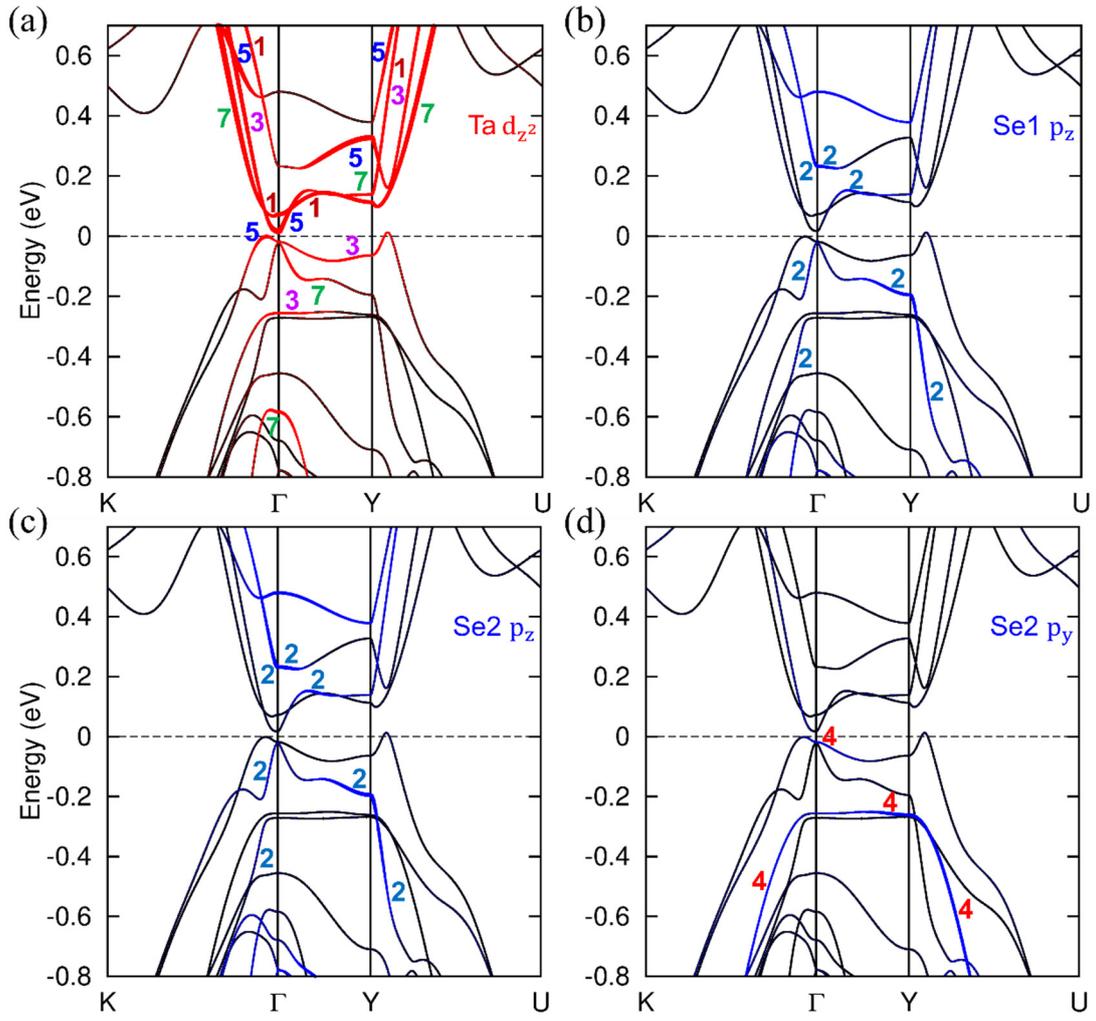

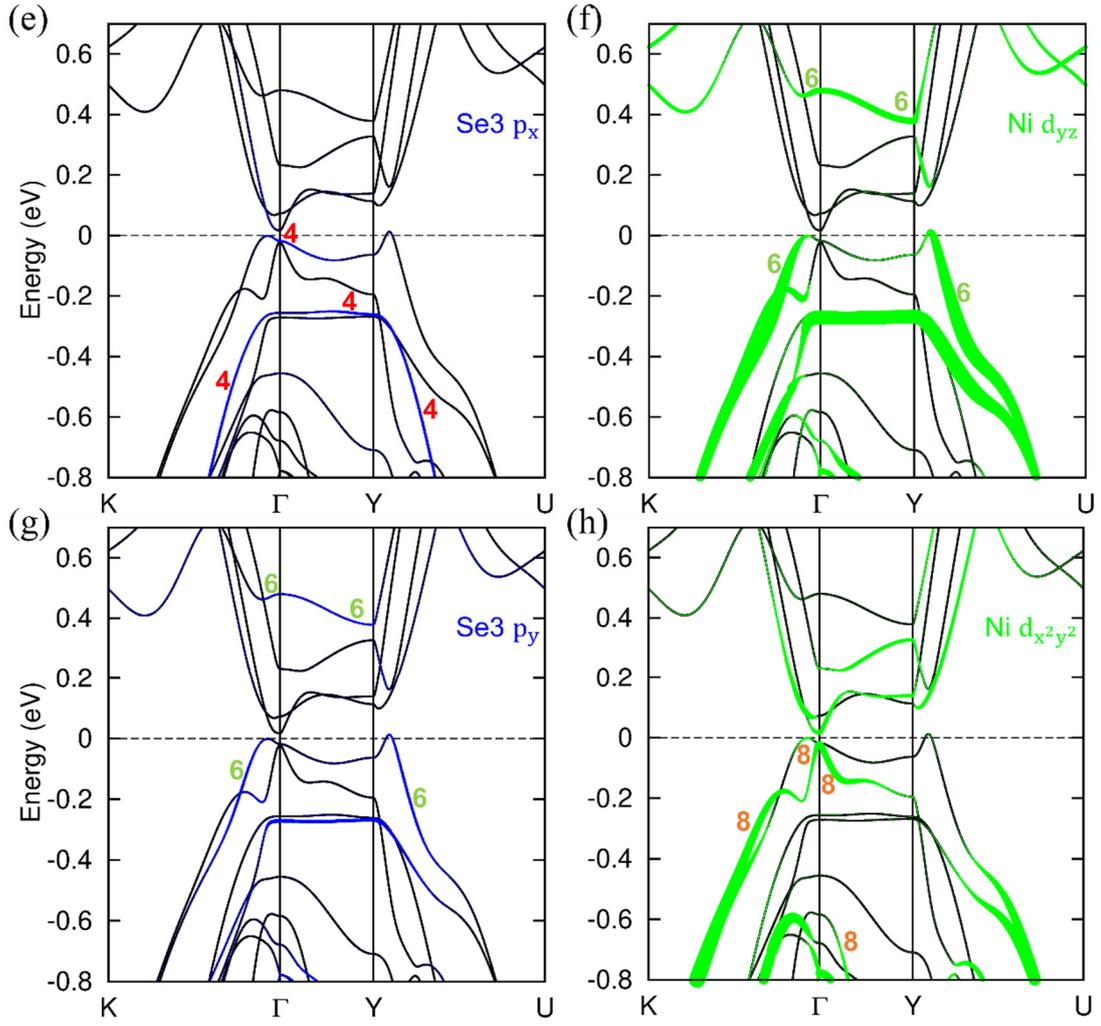

FIG. S5. Projected band structures of the monoclinic phase of Ta$_2$NiSe$_5$ without SOC (a)-(h). The sizes of the red, blue and green curves represent the weights of the Ta $d$, Ni $d$ and Se $p$ orbitals, respectively. The numbers 1-8 denote the eight bands.

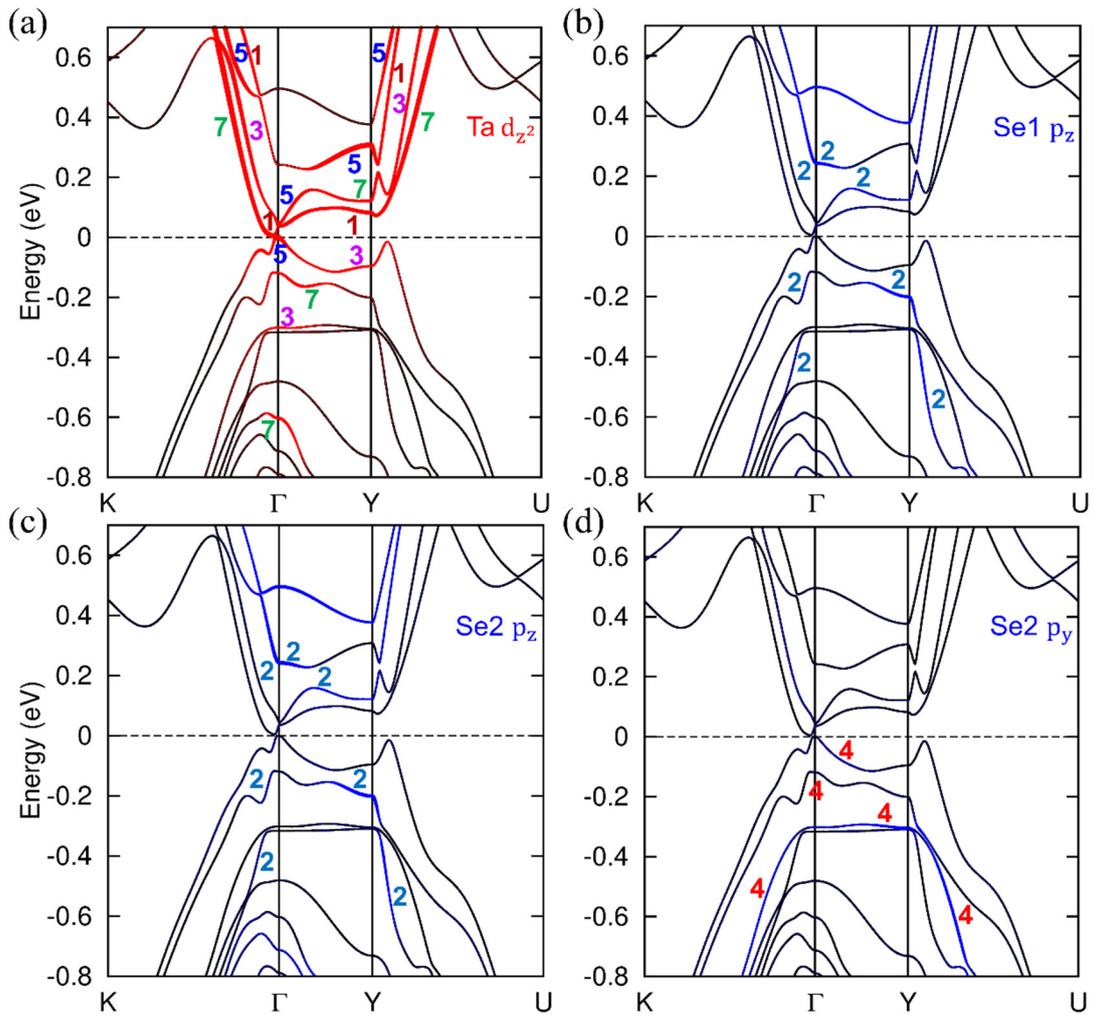

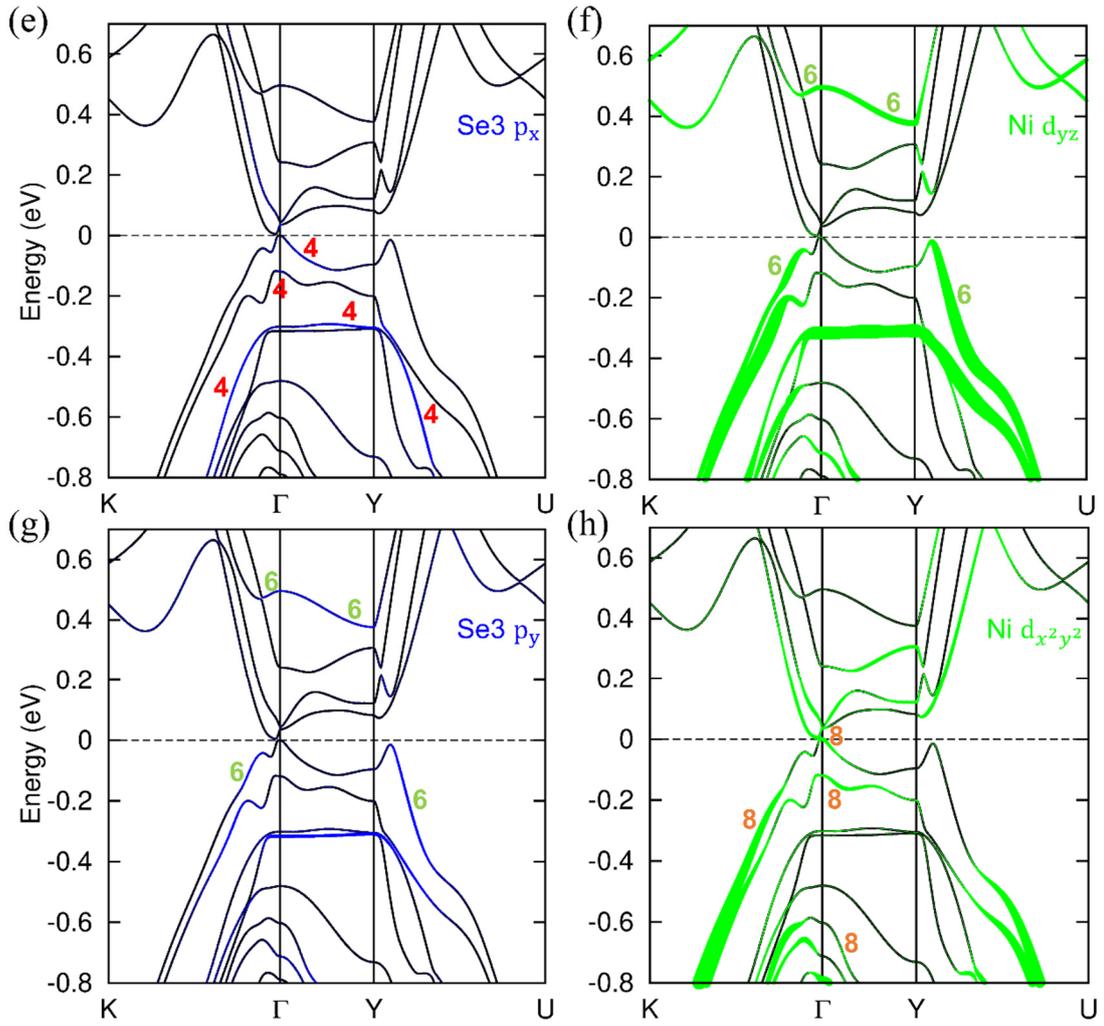

FIG. S6. Projected band structures of the monoclinic phase of Ta$_2$NiSe$_5$ with SOC (a)-(h). The sizes of the red, blue and green curves represent the weights of the Ta $d$, Ni $d$ and Se $p$ orbitals, respectively. The numbers 1-8 denote the eight bands.